\documentclass[pra,twocolumn,superscriptaddress]{revtex4-1}

\begin{document}

\title{Translating Concepts of State Transfer to Spin-1 Chains}

\author{Marcin Wie\'sniak}\affiliation{Institute of Theoretical Physics and Astrophysics, University of Gda\'nsk,\\ul. Wita Stwosza 57, 80-952 Gda\'nsk, Poland}
\author{Arijit Dutta}\affiliation{Institute of Theoretical Physics and Astrophysics, University of Gda\'nsk,\\ul. Wita Stwosza 57, 80-952 Gda\'nsk, Poland}
\author{Junghee Ryu}\affiliation{Institute of Theoretical Physics and Astrophysics, University of Gda\'nsk,\\ul. Wita Stwosza 57, 80-952 Gda\'nsk, Poland}

\begin{abstract}
State transfer is a well-known routine for various systems of spins-$\frac{1}2$. Still, it is not well studied for chains of spins of larger magnitudes. In this contribution we argue that while perfect state transfer may seem unnatural in spin-1 systems, it is still feasible for arrays of V-type three-level atoms. Tomography of such 1D array is also shown to be possible by acting on one atom from such an array.  
\end{abstract}

\maketitle

Secure cryptographic key distribution \cite{BB84} and quantum computation \cite{DEUTSCHJOZSA,SHOR,GROVER} are just two of many prospective applications of quantum information processing (QIP). While these possibilities have been intensively explored for ensembles of two-dimensional quantum systems, relatively little work has been done on higher-dimensional elementary subsystems. This seems to be so in spite of the facts that quantum cryptography with larger alphabets may be more robust against noise \cite{LARGER}, and that in quantum computing, higher dimensional systems may open the way to more efficient implementation of some protocols or realising multi-valued logics problem. We also like to stress more fundamental features of high-dimensional Hilbert spaces. For example, Kochen-Specker theorem \cite{KOCHEN} cannot be formulated for qubits, the set of all states has a far more complex structure (which is not fully still recognized).  

It is hence relevant to study more problems of utilizing higher-dimensional systems, e.g., qutrits -- with three distinc levels, as in spin-1,  in terms of their usefulness QIP. In this contribution we consider one of the most basic challenges, namely distribution of qutrit states. We want to discuss it in the fashion of transferring a state through a chain of nearest-neighbor coupled spins. This approach was suggested by Bose \cite{BOSE}. In the original proposal, a state to be transferred is initialized at one end of a chain of spins-$\frac{1}{2}$ coupled by Heisenberg or xx interaction subject to free evolutions, and the strategy is simply to wait until the fidelity of the state of the last spin to the one we uploaded is acceptably high. Such a time is predicted theoretically for the used chain. While later it was shown that Heisenberg interaction (without local magnetic fields) cannot be used to perfectly perform this task \cite{WIESNIAK} in general, for xx interaction it was noticed \cite{CHRISTANDL} that in certain subspaces the whole chain can be seen a single large spin, with inter-site coupling acting as a transverse magnetic field. The state of the chain is then rotated, leading to perfect mirroring, i.e., transfer of the information  from one end to the other. Then, more general conditions for mirroring were formulated \cite{CHINY,KAY1}, and more importantly, it was noticed that one does not need to perform additional actions, such as chain initialization \cite{PATERNOSTRO}, or even remote collaboration \cite{KAY2, MW1} to achieve perfect fidelity. Also, protocols have been proposed to attain perfect or almost perfect transfer with an arbitrary chain, both with single \cite{BUCKET} and double infrastructure \cite{DUALRAIL}. In this context, later results on tomography of such chains gain on importance. Bugrarth and Maruyama \cite{TOMO1} have shown that  coupling constants and magnetic fields of any xxz spin-$\frac{1}{2}$ chain can be estimated by acting on the first spin only, while DiFranco, Paternostro, and Kim \cite{TOMO2} demonstrated that for xx chains these parameters can be estimated without the state of the whole chain being initialized. Combination of these two methods allows to estimate the topology of nonlinear systems of spins \cite{WIESNIAK}.

Let us investigate which of these concepts can be translated to the language of spins-$1$ (or, in general arbitrary spin magnitudes). Our motivation is that, contrary to a common belief, an infinite Hilbert space dimension is not a valid classical limit. In fact, it permits stronger deviation of quantum systems from classical behavior, as mentioned above.

The problem of transferring an unknown quantum state of any dimensionality has been discussed already in, e.g., Ref. \cite{QIN}, where the authors demonstrate a high-fidelity transfer over chains of spins, each of them largely exceeding the dimension of a transferee. However, their solution uses the original approach of Bose, with non-periodic evolution, and waiting for ``the optimal time'' of transfer. In fact, this optimal time is not discussed in Ref. \cite{QIN}, and regime of a small ratio between the end-of-the chain coupling constants and the others (unmodulated) suggests it to be considerably long. The scheme is also expensive in terms of infrastructure being used -- the larger magnitude of the chain constituents, the better the fidelity. On top of that, only three and five-site chains are considered in the Reference. It is hence difficult to comment on the performance in function of the length of the chain. Here, we aim in perfect transfer of a higher-dimensional state by a minimal infrastructure. 

First, we study state transfer in chains (Heisenberg-like-coupled) of spins-1. This approach to the problem seems to be natural, as total angular momentum $L$ of multiple spins are segregated by their parity-states . This feature plays a crucial role in state mirroring, which happens in spatially symmetric systems when we can generate a $\pi$ phase difference between odd and even components of the states. Notions of parity refer here to the behavior of a state under the operation of inversion with the middle of the chain.  This requires having, up to an additive and multiplicative constants, odd eigenenergies in the odd part of the Hamiltonian and even in the even. One immediately notices that this is not possible to realize a two-site SWAP gate (which is a primitive of any perfect state transfer routine) with this interaction: even quantiplets are eigenstates of the Heisenberg interaction, $\vec{S}_1\cdot\vec{S}_2$ (single subscripts of spin operators denote the particle, double -- the particle and the direction) with eigenvalues 1, odd triplets -- with -1, and an even singlet -- with -2. However, we get the desired property if we combine the standard and the squared Heisenberg interactions,
\begin{equation}
h_{1,2}=\frac{1}{2}\left(\vec{S}_1\cdot\vec{S}_2+\left(\vec{S}_1\cdot\vec{S}_2\right)^2\right).
\end{equation}
Indeed, $e^{i\pi h_{1,2}}$ is a SWAP gate. The next quick benchmark for the possibility of building a mirroring chain is to test mirroring properties of a three-site chain,
\begin{equation}
h_{1,2,3}=h_{1,2}+h_{2,3},
\end{equation}
since a certain symmetry of the system is a key feature of all mirroring systems. This test is failed by this candidate, with
\begin{equation}
\langle 100|e^{i th_{1,2,3}}|001\rangle=\frac{1}{6}(e^{it}-3e^{3it}+2e^{4it}),
\end{equation} 
($|1\rangle$, $|0\rangle$, $|\bar{1}\rangle$ are eigenstates of $S_{\cdot,z}$ with respective eigenvalues $1,0,-1$)  
reaching only $\frac{\sqrt{3}}{4}e^{i\frac{5\pi}{6}}$ for $t=\frac{2\pi}{3}$. Still, free evolution of analogous states of four states governed by $h_{1,2}+h_{2,3}+h_{3,4}$ is irregular, which gives prospects for applying the bucket scheme. 

To find the analogue of an xx chain for spins-$\frac{1}{2}$, we consider few types of interaction. In table~\ref{table:1} we list their spectra.

\begin{table*}
\begin{tabular}{|c|c|c|c|}
\hline
Name&Form&Spectrum (even)&Spectrum (odd)\\
\hline
$O_{1}$&$S_{1,x}S_{2,x}+S_{1,y}S_{2,y}$&$\{\sqrt{2},1,1,0,0,-\sqrt{2}\}$&$\{0,-1,-1\}$\\
$O_{2}$&$S_{1,z}S_{2,z}$&$\{1,0,0,0,0,-1\}$&$\{0,0,-1\}$\\
$O_{3}$&$S_{1,x}^2S_{2,x}^2+S_{1,y}^2S_{2,y}^2$&$\{2,1,1,1,1,0\}$&$\{1,0,0\}$\\
$O_{4}$&$S_{1,z}^2S_{2,z}^2$&$\{1,1,1,0,0,0\}$&$\{1,0,0\}$\\
$O_{5}$&$S_{1,x}^2S_{2,x}+S_{1,y}^2S_{2,y}$&&\\
&$+S_{1,x}S_{2,x}^2+S_{1,y}^2S_{2,y}$&$\{\sqrt{6},\sqrt{2},0,0,-\sqrt{2},-\sqrt{6}\}$&$\{0,0,0\}$\\
\hline
\end{tabular}
\caption{Eigenvalues of chosen interaction types in odd and even subspaces.}
\label{table:1}
\end{table*} 

Table~\ref{table:1} shows that not a single of these interactions is by itself suitable for realizing perfect mirroring. One of two things happen: either we find eigenvalues of the same parity in both subspaces, or all the eigenvalues cannot be made rational at the same time. Above we have shown that specific functions of $O_{1}$ and $O_{2}$ can realize a SWAP gate for two sites, but fail for longer chains.

Since the way to realize general state transfer remains unknown, the easy solution is to limit oneself to a subspace, in which we can encode a qutrit states. Namely, we assume that the whole chain is initialized in state $|00...0\rangle$ and the interaction is able to transfer either type of excitations (up and down). Such a scheme was discussed for two ionic qutrits in Ref \cite{TR1}, and a similar solution was loosely discussed in Ref. \cite{TR2}. Notice that this partial success comes at cost: not only we need to initialize the whole system, but also it is suitable for a half-duplex communication only; if one of the users decides to send a message, the other cannot upload his message, but must wait for the delivery.

xx interaction turns out not to be suitable for this purpose. Although it can be used for mirroring states with excitations, $|00...0\rangle$ is not a stationary state of this evolution. This forces us to use the SWAP gate acting on a subspace $\sigma$ containing $|00...0\rangle$ and all states with one excitation of either type, $\{|10...0\rangle, |01...0\rangle,..., |00...1\rangle,|\bar{1}0...0\rangle,|0\bar{1}...0\rangle,...,|00...\bar{1}\rangle \}$ We introduce $S_{\cdot,u}=S_{\cdot,z}S_{\cdot,x}+S_{\cdot,x}S_{\cdot,z}$ and $S_{\cdot,v}=S_{\cdot,z}S_{\cdot,y}+S_{\cdot,y}S_{\cdot,z}$. Then
\begin{eqnarray}
A_{\cdot,1}&=|1\rangle\langle 0|=&\frac{1}{2\sqrt{2}}(S_{\cdot,u}+iS_{\cdot,v}+S_{\cdot,x}+i S_{\cdot,y}),\nonumber\\
A_{\cdot,2}&=|\bar{1}\rangle\langle 0|=&\frac{1}{2\sqrt{2}}(-S_{\cdot,u}-iS_{\cdot,v}+S_{\cdot,x}+i S_{\cdot,y}).
\end{eqnarray}
The SWAP gate is obtained by combining interaction $H_{i,j}(a,b)=a(A_{i,1}A_{j,1}^\dagger+A_{i,1}^\dagger A_{j,1})+b(A_{i,2}A_{j,2}^\dagger+A_{i,2}^\dagger A_{j,2})$ with $S_{\cdot,z}^2$. $N$-site Hamiltonian 
\begin{equation}
\label{Ham}
H=\sum_{i=1}^{N-1}H_{i,i+1}(a_i,b_i)+\sum_{i=1}^N\left(B_iS_{i,z}+C_iS_{i,z}^2\right)
\end{equation}
projected on $\sigma$ has only diagonal and next-to-diagonal entries not vanishing:
\begin{widetext}
\begin{eqnarray}
\Pi_\Sigma H\Pi_\Sigma^\dagger=\left(\begin{array}{ccccccccc}
C_1+B_1&a_1&0&...&0&&&&\\
a_1&C_2+B_2&a_2&...&0&&&&\\
0&a_2&C_2+B_3&...&0&&&&\\
...&...&...&...&...&&&&\\
0&0&0&...&0&0&0&0&...\\
&&&&0&C_1-B_1&b_1&0&...\\
&&&&0&b_1&C_2-B_2&b_2&...\\
&&&&0&0&b_2&C_2-B_3&...\\
&&&&...&...&...&...
\end{array}\right)
\end{eqnarray}
\end{widetext}
with states ordered as follows:$|10..0\rangle, |01...0\rangle,...,|00...0\rangle,|\bar{1}0...0\rangle, |0\bar{1}...0\rangle,...,|00...\bar{1}\rangle$. Now, by choosing proper coupling constants, we can satisfy conditions for perfect mirroring described in Refs. \cite{CHINY,KAY1} and, in particular, adopt values from Ref. \cite{CHRISTANDL}. The diagonal terms are to fix the difference between the eigenvalue of $|00...0\rangle$ and of all other states. Since the former is $0$, we expect to have only odd eigenvalues for odd states and even for even ones. For choice $a_i=b_i=\frac{\sqrt{i(N-i)}}2$, we shall choose $B_i=0$ and $C_i=\frac{N}{2}$.

Finally, we pass to the problem of chain tomography despite limited access. We briefly recall that in Refs. \cite{TOMO1,TOMO2,TOMO3} the structure of a chain was concluded from probabilities of the revival of the initial state, or of its transition to another state, measured at short times.
Since transfer requires chain initialization, it is natural to adopt the technique from Ref. \cite{TOMO3}.
Namely, we first initialize the unknown chain in state $|10...0\rangle$, then at various time we measure the first spin in the computational basis, $\{|1\rangle,|0\rangle,|\bar{1}\rangle\}$. From recurrence probability  we can conclude eigenergies of the Hamiltonian eigenstates with one up excitation and their overlaps with with $|10...0\rangle$. As shown in Ref. \cite{TOMO3}, by solving the eigenproblem of $H$ we can conclude $|a_{i,i+1}|$ and $B_i+C_i$. A similar procedure for $|\bar{1}\rangle$ gives $|b_{i,i+1}|$ and $B_i-C_i$. The signs of the coupling constants must be known by assumption. 

In conclusions, we have investigated which concepts known from spin-$\frac{1}{2}$ chains are suitable for transferring an unknown quantum state through a spin-$1$ chain. This problem might become particularly relevant for manipulating arrays of three-level atoms. We have argued that the most natural inter-spin coupling types cannot perform this task perfectly by their own. Remarkably, while some interactions realize the SWAP gate between two spins-1, they fail in this task for three subsystems, which disqualifies them as candidates for longer chains. However, it is possible to transfer any qutrit state through an arbitrarily long chain, using coupling constants known from the studies of spin-$\frac{1}{2}$ chains, and some artificial interaction, which realizes SWAP gates in two-dimensional subspaces. This is equivalent of the xx coupling for spins-$\frac{1}{2}$, but in contrast it requires state initialiaztion of the whole chain. A suitable V-type structure of energetic levels can be found in, e.g., Rubidium atoms, which would decay to a ground state in low temperatures, Since there is no exchange mechanism between $|1\rangle$ and $|\bar{1}\rangle$, the chain can be used only in a simplex mode, without the possibility of sending two messages in the opposite directions at the same time. By using a routine for spin-$\frac{1}{2}$ chains, we can also perform tomography of the chain (estimate the coupling constants, magnetic field magnitudes and $C_i$s).
 
Notice that our results shows that since propagation of an initial state is still possible, one can use more elaborate techniques to retrieve the initial state at the other chain with the bucket \cite{BUCKET}. While in the spin-$1$ formalism the interaction discussed here is very artificial, it seems natural in arrays of three-level atoms with one of the transitions strongly forbidden. Indeed, V-type spectrum is very suitable for these purposes, as in low temperatures all elements of our chain would thermalize to a state, which we would call $|0\rangle$ as a convention.  

M. W. is supported by IDEAS PLUS grant (IdP2011 000361) and program HOMING PLUS from the Foundation for Polish Science. A. D. is supported by MPD project by the Foundation for Polish Science and J. R. acknowledges support from the Foundation for Polish Science TEAM project cofinanced by the EU European Regional Development Fund and a NCBiR-CHIST-ERA Project QUASAR.

\end{document}